\documentclass[aps,prl,twocolumn,superscriptaddress,floatfix]{revtex4-1}

\usepackage{graphics}
\usepackage{epsfig}
\newcommand{\abu}
{\affiliation{Department of Physics, Boston University, Boston, MA
02215, USA}}

\newcommand{\auiuc}
{\affiliation{Department of Physics, University of Illinois at
Urbana-Champaign, Urbana, IL 61801, USA}}
\newcommand{\ajmu}
{\affiliation{Department of Physics, James Madison University,
Harrisonburg, VA 22807, USA}}
\newcommand{\aky}
{\affiliation{Department of Physics and Astronomy, University of
Kentucky, Lexington, KY 40506, USA}}
\newcommand{\akvi}
{\affiliation{KVI, University of Groningen, NL-9747AA Groningen, The
Netherlands}}
\newcommand{\akwu}
{\affiliation{Department of Mathematics and Physics, Kentucky
Wesleyan College, Owensboro, KY 42301, USA }}
\newcommand{\aregis}
{\affiliation{Department of Physics and Computational Science, Regis
University, Denver, CO 80221, USA}}
\newcommand{\apsi}
{\affiliation{Paul Scherrer Institute, CH-5232 Villigen PSI,
Switzerland}}

\begin{document}

\title{Measurement of the Positive Muon Lifetime and Determination
of the Fermi Constant to Part-per-Million Precision}

\author{D.M.\,Webber}
\auiuc
\author{V.\,Tishchenko}
\aky
\author{Q.~Peng}
\abu
\author{S.\,Battu}
\aky
\author{R.M.~Carey}
\abu
\author{D.B.\,Chitwood}
\auiuc
\author{J.\,Crnkovic}
\auiuc
\author{P.T.\,Debevec}
\auiuc
\author{S.\,Dhamija}
\aky
\author{W.\,Earle}
\abu
\author{A.\,Gafarov}
\abu
\author{K.\,Giovanetti}
\ajmu
\author{T.P.\,Gorringe}
\aky
\author{F.E.\,Gray}
\aregis
\author{Z.\,Hartwig}
\abu
\author{D.W.\,Hertzog}
\auiuc
\author{B.\,Johnson}
\akwu
\author{P.\,Kammel}
\auiuc
\author{B.\,Kiburg}
\auiuc
\author{S.\,Kizilgul}
\auiuc
\author{J.\,Kunkle}
\auiuc
\author{B.\,Lauss}
\apsi
\author{I.\,Logashenko}
\abu
\author{K.R.\,Lynch}
\abu
\author{R.\,McNabb}
\auiuc
\author{J.P.\,Miller}
\abu
\author{F.\,Mulhauser}
\auiuc \apsi
\author{C.J.G.~Onderwater}
\auiuc \akvi
\author{J. Phillips}
\abu
\author{S.\,Rath}
\aky
\author{B.L.\,Roberts}
\abu
\author{P.\,Winter}
\auiuc
\author{B.\,Wolfe}
\auiuc

\collaboration{MuLan Collaboration}

\begin{abstract}
We report a measurement of the positive muon lifetime to a precision
of 1.0~parts per million (ppm); it is the most precise particle
lifetime ever measured. The experiment used a time-structured,
low-energy muon beam and a segmented plastic scintillator array to
record more than $2 \times 10^{12}$ decays. Two different stopping
target configurations were employed in independent data-taking
periods. The combined results
give $\tau_{\mu^+}(\rm{MuLan}) = 2196980.3(2.2)$~ps, more than 15
times as precise as any previous experiment. The muon lifetime gives
the most precise value for the Fermi constant: $G_F(\rm{MuLan}) =
1.1663788 (7)\times 10^{-5}$~GeV$^{-2}$ (0.6~ppm).  It is also used
to extract the $\mu^{-}p$ singlet capture rate, which determines the
proton's weak induced pseudoscalar coupling ${\textsl g}^{}_P$.

\end{abstract}
\maketitle

A measurement of the positive muon lifetime, $\tau_{\mu^+}$, to high
precision determines the Fermi constant, $G_{F}$, according to the
relation
\begin{equation}
\frac{1}{\tau_\mu}= \frac{G_{F}^2 m_\mu^5}{192 \pi^3} \left(
1+\Delta q \right). \label{eq:muondecay}
\end{equation}
Here, $\Delta q$ represents well-known phase space and both QED and
hadronic radiative corrections~\cite{oldQED}, and we assume that
$G_F$ is universal for weak interactions. Strictly speaking,
$\tau_{\mu^+}$ determines a muon-decay-specific coupling, denoted
$G_{\mu}$, which could be compared to other $G_{F}$ determinations
as a test of the standard model~\cite{fermiconstants}.

Prior to 1999, the limitation on the precision of $G_F$ was
dominated by the uncertainty on $\Delta q$. Van Ritbergen and Stuart
were the first to compute the second-order QED radiative corrections
in the massless electron limit, reducing the theoretical uncertainty
to below 0.3~ppm~\cite{vanRitbergen:All}, and well below the
then-current experimental uncertainty. This development motivated a
new generation of precision muon lifetime measurements,
MuLan~\cite{MuLan:2007} and FAST~\cite{FAST:2008}. More recently,
Pak and Czarnecki extended the result in \cite{vanRitbergen:All} to
finite electron mass, which shifts the predicted decay rate
$1/\tau_\mu$ by -0.43~ppm; alternatively, it decreases $G_F$ by
0.21~ppm~\cite{Pak}.

In Ref.~\cite{MuLan:2007}, we reported an 11~ppm measurement of
$\tau_{\mu^+}$ based on a relatively short
commissioning run. This Letter reports the results from a 100 times
larger data set, accumulated using the final setup of the experiment.

The experiment is designed to stop muons
in a target during a beam-on accumulation interval and measure the
decay positrons---primarily from
the $\mu^+ \rightarrow e^+ \nu_e \bar{\nu}_\mu$ decay mode---during
a beam-off measurement period.
The two
running periods, R06 and R07, used different targets.
More than $10^{12}$ decays were recorded in each
period.


The experiment used the $\pi$E3 beamline at the Paul Scherrer
Institute (PSI). During the run, positive muons from at-rest pion
decay near the surface of the production target are directed to the
experiment through two opposing vertical dipole magnets and a series
of 15 magnetic quadrupole lenses. A velocity-selecting
$\vec{E}\times\vec{B}$ separator is tuned to pass muons and reject
positrons.
A special feature of the beamline is a custom, 60-ns switching,
25-kV kicker~\cite{kicker}. When energized, the electric field
across the 120-mm vertical gap and 1500-mm length displaces the muon
beam by 46~mm at the exit and deflects it by 45~mrad onto a
downstream collimator.  The muon flux of $\sim10^{7}~\mu$/s at the
experimental focus is reduced by an extinction factor $\epsilon
\sim10^3$. The kicker is switched by an external timing circuit that
synchronizes the data collection cycles into $5~\mu$s kicker-off
accumulation periods, $T_A$, followed by $22~\mu$s kicker-on
measurement periods, $T_M$, also called ``fills.'' The kicker
voltage stability during the measuring period, $V(t)$, coupled with
the derivative $d\epsilon /dV(t)$ at full voltage, determines the
time stability of background from this source.


Muons which arrive at the target are nearly $100\%$ polarized.
Because of parity violation in the decay, positrons are emitted
preferentially in the direction of the muon spin.  Any change in the
ensemble-averaged residual polarization $\langle
\overrightarrow{P_{\mu}}\rangle$ during the measuring period can
result in an effective time-dependent acceptance if the individual
detectors are not symmetric around the stopping target and identical
in response. For many materials, the magnitude of $\langle
\overrightarrow{P_{\mu}}\rangle$  is reduced during the muon
stopping process. It can also be reduced by the application of a
magnetic field transverse to the spin direction and having a field
strength sufficient to precess the spin rapidly compared to $T_A$.
In practice, a residual longitudinal polarization $\langle
\overrightarrow{P_L}\rangle$ remains, owing to misalignment of the
transverse field. The array of detectors is highly and symmetrically
segmented. The sum of events recorded by a detector at angle
$(\theta, \phi)$ with those from an equally efficient detector at
$(180^\circ - \theta, 180^{\circ}+\phi)$ form a decay histogram that
is immune to changes in $\langle \overrightarrow{P_{\mu}}\rangle$.

\begin{figure}
\begin{center}
\includegraphics[width=\columnwidth]{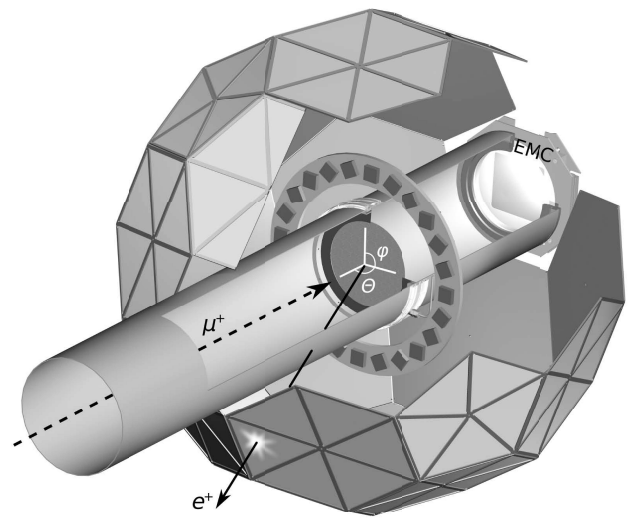}
\end{center}
\caption{Diagram of the experiment. Muons are transported to the
AK-3 or quartz disk targets through a vacuum beampipe, which is
lined on the inside with 0.1-mm-thick AK-3 foils polarized in the
azimuthal direction. 170 pairs of triangular scintillator detectors,
each read out individually, surround the target. The Halbach
magnetic ring is used only for the quartz disk. A wire chamber (EMC)
is placed at the rear to monitor the beam profile when the targets
are opened. } \label{fig:detector}
\end{figure}

Several changes compared to the setup used in Ref.~\cite{MuLan:2007}
involve the muon delivery and the stopping target.  A
200-mm-diameter vacuum beampipe now extends the beamline through the
detector array (see Fig.~\ref{fig:detector}) ending in a thin mylar
window. Muons are stopped in a 21-mm horizontal by 10-mm vertical
(RMS) spot centered on the target disk. The R06 target is a
0.5-mm-thick, 200-mm-diameter ferromagnetic AK-3
foil~\cite{arnokrome}, having an internal in-plane bulk magnetic
field of approximately 0.4~T.  The R07 target is a 2-mm-thick,
130-mm-diameter crystalline quartz (SiO$_2$) disk, in which stopped
muons form muonium $90\%$ of the time~\cite{brewer}. A Halbach
arrangement of permanent magnets provides a nearly uniform $0.013$~T
field in the plane of the quartz disk. The internal AK-3
magnetization or the external magnet field normally points left or
right in the horizontal direction and the muon spin precesses in the
vertical plane with periods of 18~ns and 2.6~ns, respectively, for
free muons in AK-3 or muonium in quartz. The $10\%$ of muons in
diamagnetic states in quartz precess with a period of $550$~ns,
which is observed. Both targets can swing open to allow the beam to
pass unobstructed to a proportional wire chamber, which monitors the
rate and transverse profile at the downstream end of the experiment.

The detector array consists of 170 stacked pairs of 3-mm-thick plastic
scintillators.  They are arranged in a truncated icosahedron geometry
and grouped in 20 hexagon and 10 pentagon houses. The center of a
pair of detectors is 383~mm from the target. Each scintillator is viewed by
either a Photonis (PH) or Electron Tubes (ET) 29-mm photomultiplier
(PMT). On average, 80~photoelectrons are registered for each
minimum-ionizing particle (MIP), which yields a well-defined pulse
shape from through-going decay positrons. The pulse-height
distribution fit to a Landau function convoluted with a Gaussian
response provides a calibration point for each detector. The most
probable value (MPV) is used as a relative calibration of the gain
stability versus time in the experimental run and, most importantly,
versus time during the measuring cycle. An LN203 nitrogen laser
distributes 300-ps-long UV pulses (337~nm) to 24 of the detectors
and to an externally located reference counter; this pulse excites
the scintillator similarly to an ionizing particle. The asynchronous
pulsing of the laser provides a sharply timed common pulse that
allows for a relative time stability monitor during the measurement
period and an independent confirmation of gain stability.

The signals from the 340 PMTs are recorded using $450$~MHz, 8-bit,
waveform digitizers (WFDs). Each has four independent inputs
associated with the inner/outer elements of a pair of detectors and the
geometrically opposite pair. The WFD sampling frequency is set
by an Agilent E4400 signal generator having stability better than
0.01~ppm/month. The frequency was set to $\pm 450$~ppm of 451~MHz,
with the exact value unknown to the collaboration and only disclosed
after the completion of a blinded analysis of the two datasets. The
``clock tick'' (c.t.) time units have the conversion 1~c.t. $\approx
2.2$~ns. Normally 24 ADC samples ($53$~ns) comprise a full waveform;
they are recorded when an input signal exceeds a set threshold, with
four to eight samples preceding the trigger point. The digitization
is extended if the threshold is exceeded at the 24th sample. For the
average pulse, the full width at $20\%$ maximum is 9~ns.


The 130~TB of raw waveforms are first converted into lists of valid
``hits'' having characteristic times and energies for each detector.
The waveforms are fit to pulse-shape templates, prepared from a set
of low-rate events; the time of a decay is defined as the peak of
the pulse shape. In more than $99.9\%$ of the cases, a single pulse
exists on a waveform and it can be identified reliably.  A
distribution of pulse energies is made for each detector and a
run-by-run dependent pulse threshold is established at the minimum
in the distribution between the low-amplitude PMT noise and the MIP
peak ($45 - 55\%$ of the MIP peak amplitude). Rare multiple-pulse
waveforms are more difficult to characterize, requiring an iterative
approach that adds additional pulses as needed and/or introduces
more sophisticated fitting methods. When two pulses are separated by
more than 3~c.t., the procedure works well. For shorter separation
times, only one hit is reconstructed and its time is set to the
energy-weighted average of the ADC samples. For resolved pulses, an
``artificial deadtime'' (ADT) can still be applied on a per-detector
basis, eliminating pulses when a minimum time separates sequential
hits in the same scintillator. The ADT used in the analysis varies from 5 to
$68~{\rm c.t.}$ and represents an important diagnostic of the pileup
correction procedure.

Histograms are filled with inner-outer coincident events for each detector pair;
the coincidence window width is set equal to the ADT.
The muon lifetime is obtained from a fit to the sum of the
170 individual histograms using the function
\begin{equation}
F(t) = A\exp(-t/\tau_\mu) + B, \label{eq:fit}
\end{equation}
where $B$ accounts for the flat background. Under our running
conditions, uncorrected pileup would shift the fitted lifetime from its
true value by $\delta\tau_\mu/$ADT = 24~ppm/c.t.  Pileup is
corrected using a statistical procedure based on the data itself. When a hit is
observed at time $t_i$, in fill $j$, an interval between $t_i$ and
$t_i +$~ADT is searched in fill $j+1$. If a hit is observed in this
interval, its time is recorded in a separate histogram, which is
then added back into the original decay histogram. The process is
repeated for higher-order pileup, and it includes fine adjustments
to account for time jitter between inner-outer coincidences, and for
uncorrelated single hits and other distortions.  A Monte-Carlo study
verified the pileup correction procedure with input hits occurring
at actual coincidence, singles, and background rates, and with the
measured time jitter in inner-outer hits.  It was also tested at
much higher rates where the correction is correspondingly larger.
Pileup-related lifetime shifts are reduced to
$\delta\tau_{\mu}/$ADT~=~+0.018~ppm/c.t. A linear extrapolation to
zero ADT gives our quoted value. The overall pileup systematic
uncertainty is estimated to be less than $0.2$~ppm.  An alternative approach
is to pileup correction is to add an explicit 1st-order pileup term, $\exp(-2t/\tau_\mu)$, to the fit function in Eq.~\ref{eq:fit}.  This yields the same lifetime as our correction procedure,
but at the price of doubling the
uncertainty on $\tau_\mu$, owing to the correlation of the additional
term in the fit.

Because the residual polarization in the AK-3 target is small enough
that it is not distinguishable in individual time spectra,
Eq.~\ref{eq:fit} describes the R06 data well. A fit to the
pileup-corrected event histogram, summed over all detector pairs, gives
a $\chi^2/{\rm dof} = 1.03 \pm 0.04$ for a fit start time $220$~ns
into the measurement period and an ADT of 6~c.t. Although no $\phi$
dependence is observed for individual fits, a 12~ppm slope
exists in the distribution of lifetime versus $\cos(\theta)$, which is canceled to
better than 0.10~ppm by the symmetry of the detector array and the
accuracy of the alignment. The slope is consistent with the
relaxation of a spin component along the beam axis. The lifetimes of
the 85 opposite detector pairs, which are first summed, then fit, is
flat, when plotted from $\theta = 0$ to 90 degrees, and the
distribution of results is Gaussian having a normalized width
$\sigma / {\rm mean} = 0.90 \pm 0.08$, which is in acceptable agreement with 1.0.

In contrast, the R07 quartz data are complicated by the relatively
slow precession and relaxation of the diamagnetic muons. The
amplitude of the precession signal for individual detectors can be as
large as $0.1\%$ and the longitudinal polarization can be as large
as $0.15\%$. A different analysis procedure, which is insensitive to
small and unknown efficiency differences, is used. Each
detector-pair histogram is first fit using Eq.~\ref{eq:fit} with $A$
multiplied by $[1+P_2\cdot\exp{(-t/\tau_2)}\sin(\omega t+\phi)]$;
here $P_2$ and $\tau_2$ are the transverse polarization and
relaxation time constant, respectively. These fits result in 170
``effective'' muon disappearance rates, which differ from the true
muon decay rate because of the relaxation of the longitudinal spin
component. Next, the 170 effective lifetimes are fit to extract the
true muon lifetime using a function that accounts for the position of
detectors relative to the longitudinal polarization direction.
This fit takes advantage of the smooth geometrical-based evolution of the relaxation effect.
The leading systematic uncertainty of 0.20~ppm for this procedure is due
to the uncertainty of the beam position on the target. The lifetime
obtained is in agreement (to 0.3~ppm) with the result from a fit to the simple sum of
detector-pair histograms using Eq.~\ref{eq:fit}.

Instability of the combined detector and electronics response, or
``gain,'' during the measurement period is a possible source for a
systematic error. It will result in gaining or losing hits, which
can shift the fitted lifetime. Four effects were identified: (i) When
triggered to begin the recording of fills, a fanout unit
induced a $1.3 \times 10^{-4}$ gain oscillation in each WFD. The
effect, which is reproducible in bench tests, disappears in
$\sim4~\mu$s. An empirical function is used to characterize and
correct for the oscillation prior to the final fits. (ii) A
short-time-scale gain shift occurs when one hit closely follows a
preceding hit. The second hit is affected by the remnant
long-decay-time scintillator light from the first pulse and by the
intrinsic recovery time of the PMT and voltage divider. A
$\sim1\%$ increase in detector gain exists following a pulse, which
persists for $50$ or $500$~ns in PH- or ET-instrumented detectors,
respectively. (iii) A long-time-scale change also exists that
differs by PMT type. The PH-tube gains increase by $1.8 \times
10^{-4}$ and the ET-tube gains decrease by $5 \times 10^{-4}$, over
the 22~$\mu$s measurement period. (iv) When below-threshold pulses
fall either in the peak or pedestal regions of a trigger pulse, they
can either raise or lower the trigger pulse amplitude, respectively.
Strictly speaking, this is a pileup effect, which is not accounted
for in the pileup-correction procedure.

These four phenomena are interrelated and give rise to an overall
variation of $3 \times 10^{-4}$ in the gain during the fill. This
variation was separately evaluated for PH- and ET-instrumented
detectors using the MPV versus time-in-fill, and was used, after
normalization and prior to the final lifetime fits,
to correct for the resulting time spectrum
distortion. After all corrections are applied, the shift in lifetime
compared to uncorrected spectra is $+0.50 \pm
0.25$~ppm.  The procedure was tested by accumulating a
high-threshold spectrum, which amplifies the time spectrum
distortion by roughly a factor of 60. The lifetimes derived from
normal- and high-threshold settings, following the correction
procedures, are in good agreement.

The complete summary of systematic uncertainties is given in
Table~\ref{tbl:ErrorTable}, including small effects related to muon
stops upstream of the target, the short-term and long-term time
response stability based on the laser system, and the uncertainty on
the absolute clock frequency.

\begin{table}[h!]
  \centering
  \caption{Systematic and statistical uncertainties in ppm.
  The errors in different rows of the table are not correlated to
  each other.  If only one error appears in a given row, the effect is $100\%$
  correlated between the two run periods.}
  \label{tbl:ErrorTable}
  \begin{tabular}{lcc}
    \hline
    Effect uncertainty in ppm       & ~~R06~~       &  ~~R07~~ \\
    \hline\hline
    Kicker stability                & ~~0.20~~  & ~~0.07~~ \\
    Spin precession / relaxation~~~~~    & ~~0.10~~  & ~~0.20~~ \\
    Pileup                          & \multicolumn{2}{c}{0.20} \\
    Gain stability                   & \multicolumn{2}{c}{0.25} \\
    Upstream muon stops             & \multicolumn{2}{c}{0.10} \\
    Timing stability                & \multicolumn{2}{c}{0.12} \\
    Clock calibration               & \multicolumn{2}{c}{0.03} \\
    \hline
    Total systematic                & 0.42 & 0.42 \\
    \hline
    Statistical uncertainty         & 1.14 & 1.68  \\
    \hline
    \hline
  \end{tabular}
\end{table}

The stability of the result versus start time of the fit is a
powerful collective diagnostic because pileup, gain stability and
spin effects all might exhibit time dependence. For both R06 and
R07, the lifetime does not depend on the fit start time, apart from
the statistically allowed variation. Further, it does not depend on
run number or magnetic field orientation.

\begin{figure}
\begin{center}
\includegraphics[width=\columnwidth]{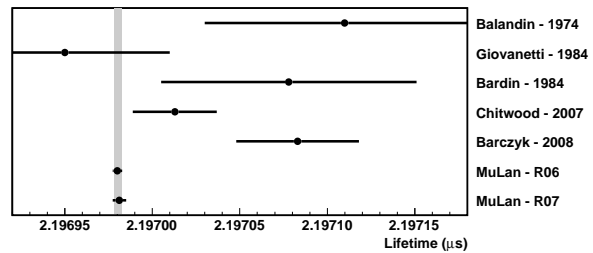}
\end{center}
\caption{Lifetime measurement summary.  The MuLan R06 and R07
results are plotted separately to illustrate the consistency. The
vertical shaded band is centered on the MuLan weighted average with
a width equal to the combined uncertainty.} \label{fig:history}
\end{figure}

The final results for the two running periods are in excellent
agreement:
\begin{eqnarray}\nonumber
\tau_{\mu}({\rm R06})&=& 2196979.9 \pm 2.5 \pm 0.9 {\rm ~ps}, \\
\tau_{\mu}({\rm R07})&=& 2196981.2 \pm 3.7 \pm 0.9 {\rm ~ps}.
\end{eqnarray}
Here, the first errors are statistical and the second systematic.
The comparison between R06 and R07 affirms, at the ppm level, the
expectation that the lifetime of bound muonium does not differ
appreciably from the free lifetime~\cite{Marciano:2000}.
 Combined we obtain
\begin{equation}
\tau_{\mu}({\rm MuLan}) = 2196980.3 \pm 2.2 {\rm ~ps~~~(1.0~ppm)},
\label{finalresult}
\end{equation}
which is in agreement with our previous
measurement~\cite{MuLan:2007}.  The error is the quadrature average
of statistical and systematic errors, where the full error matrix calculation,
including all correlations, is used to combine uncertainties.
The MuLan result is more than 15
times as precise as any other individual
measurement~\cite{FAST:2008,lifetimes} and consequently dominates
the world average. Our result lies $2.5~\sigma$ below the current
PDG average~\cite{PDG:2008}. Figure~\ref{fig:history} shows the
recent history of measurements, together with the MuLan average.

Our value for $\tau_{\mu^+}$ leads to the most precise determination
of the Fermi constant:
\begin{equation}
G_F({\rm MuLan}) = 1.1663788(7) \times 10^{-5} {\rm
~GeV^{-2}}~~~({\rm 0.6~ppm}).
\end{equation}
The positive muon lifetime is also used to obtain ordinary muon
capture rates in hydrogen~\cite{MuCap:2007} or
deuterium~\cite{MuSun:proposal} by the lifetime difference method,
$\Gamma_{{\rm cap}} = 1/\tau_{\mu^-} - 1/\tau_{\mu^+}$. These
capture rates determine hadronic quantities as discussed
in~\cite{PeterReview}. For example, our new result lowers the
$\mu^{-}p$ capture rate used in Ref.~\cite{MuCap:2007} by
8~s$^{-1}$, and thus shifts ${\textsl g}^{}_P$ upward to even better
agreement with theory. Finally, the improved precision reduces the
$\tau_{\mu^+}$ uncertainty in the determination of muon capture in
\cite{MuCap:2007} and future efforts to below 0.5~s$^{-1}$.

We thank the PSI staff, especially D.~Renker, K.~Deiters,
and M. Hildebrandt; M. Barnes and G. Wait from TRIUMF for the kicker design, the NCSA for enabling
and supporting the data analysis effort, and the U.S. National Science Foundation
for their financial support.

\end{document}